\documentclass[12pt]{iopart}

\usepackage{iopams}  
\usepackage{graphicx}
\usepackage{subfig}
\usepackage[breaklinks=true,colorlinks=true,linkcolor=blue,urlcolor=blue,citecolor=blue]{hyperref}
\usepackage{romannum}
\usepackage{upgreek}
\expandafter\let\csname equation*\endcsname\relax
\expandafter\let\csname endequation*\endcsname\relax
\usepackage{amsmath}
\usepackage{physics}
\usepackage{xcolor, soul}
\usepackage[normalem]{ulem}
\begin{document}

\title[Rubidium atom spectral lineshapes in high intensity electric fields]{Rubidium atom spectral lineshapes in high intensity electric fields near an optical nanofibre}

\author{Vandna Gokhroo$^1$, Fam Le Kien$^1$, S\'ile {Nic Chormaic}$^{1,2}$}

\address{$^1$Okinawa Institute of Science and Technology Graduate University, Onna, Okinawa 904-0495, Japan \\
$^2$Universit\'e Paris-Saclay, CNRS, Laboratoire de Physique des Gaz et des Plasmas, 91405 Orsay, France}
\ead{sile.nicchormaic@oist.jp}
\vspace{10pt}
\begin{indented}
\item[]January 16, 2022
\end{indented}

\begin{abstract}
The integration of cold atomic systems with optical nanofibres is an increasingly important experimental platform. Here, we report on the  spectra observed during a strongly driven, single-frequency, two-photon excitation of cold rubidium atoms near an optical nanofibre. At resonance, two competitive processes, namely a higher excitation rate and  stronger pushing of atoms from the nanofibre due to resonance scattering, need to be considered. We discuss the processes that lead to the observed two-peak profile in the fluorescence spectrum as the excitation laser is scanned across the resonance, noting that the presence of the optical nanofibre dramatically changes the fluorescence signal. These observations are useful for  experiments where high electric field intensities near an ONF are needed, for example when driving nonlinear processes.    
 \end{abstract}

%
%
%
%
%

\section{Introduction:}

Interactions between  atoms and laser fields have been widely studied in many different scenarios, such as a two-level atom interacting with a monochromatic laser field, a three-level atom interacting with two laser fields, multilevel atoms with multifrequency light fields \cite{1998atomPhotonCohen, agarwal_2012}, etc. The presence of coherence in such systems can substantially enhance the optical nonlinearity, leading to many interesting effects such as electromagnetically induced transparency (EIT) and coherent population trapping \cite{RevModPhys.Fleischhauer, PhysRevA.Fulton, Kumar_2015}. Over the past  decade, optical waveguides, including optical nanofibres (ONF), have come to the fore as an exciting platform to study interactions of atoms with light. We focus our discussion on optical nanofibres, which provide tight confinement of light beyond the Rayleigh range and also serve as an ideal channel for light propagation. Optical nanofibres have been used to efficiently probe, manipulate and trap cold, alkali atoms \cite{goban_demonstration_2012,lee_inhomogeneous_2015,Nieddu_2016, two_nanofibers_2021, gupta2021machine} with recent works extending atom and ONF studies to the observation of a quadrupole transition in $^{87}$Rb atoms \cite{2020quadrupole}, the demonstration of  quadrature squeezing of nanofibre-guided light \cite{PhysRevLett.127.123602}, and a determination of the polarization dependency of spin selection in laser-cooled atoms \cite{PhysRevResearch.2.033341}.   

One particular advantage   of ONFs is that they provide a means to perform high electric field intensity experiments at very low optical powers.  However, there is also a disadvantage since the origin of some observed effects on spectral lineshapes due to the presence of the ONF is not always clear \cite{PhysRevA.72.032509, PhysRevLett.99.163602, Nayak:07, Nayak2008split}.   In this work, we generate high electric field intensities using either an optical nanofibre or a tightly focussed free-space beam, so as to make a comparison between the observed effects on spectral line shapes. Notably, the fluorescence signal from  atoms near an ONF can exhibit a two-peak profile as the frequency of the excitation laser is scanned across an atomic resonance.   We explore several atomic transitions and experimental configurations to investigate this effect in detail and hypothesise on possible sources of the observed phenomenon.
 
\section{Experiment}

\subsection{Experimental setup}
We prepared an ensemble of trapped cold $^{87}$Rb atoms using a standard magneto-optical trap (MOT) configuration \cite{PhysRevResearch.2.033341}. The average temperature of the atomic cloud was $\sim$~200~$\mu$K, with a density of 10$^{10}$ atoms/cm$^3$. The cooling beams for the MOT were  14~MHz red-detuned from the 5S$_{1/2}$ F=2 $\rightarrow$ 5P$_{3/2}$ F$'$=3 transition with an intensity of $\sim$11~mW/cm$^2$ per beam. The optical nanofibre was fabricated from commercial step-index optical fibre (Fibercore, SM800-5.6-125) by heating it with an oxygen-hydrogen flame and simultaneously pulling it using computer-controlled motorised stages \cite{doi:10.1063/1.4901098}. The final nanofibre waist was $\sim$~400~nm, allowing only single-mode propagation for all the wavelengths used in the experiment, 780 nm, 795 nm,  and 993 nm. Note that all optical nanofibre-guided excitation laser powers mentioned were measured at the output pigtail of the ONF and the power at the waist could be higher due to propagation losses in the fiber.

\begin{figure}[!tbp]
  \centering
   \includegraphics[width=\textwidth]{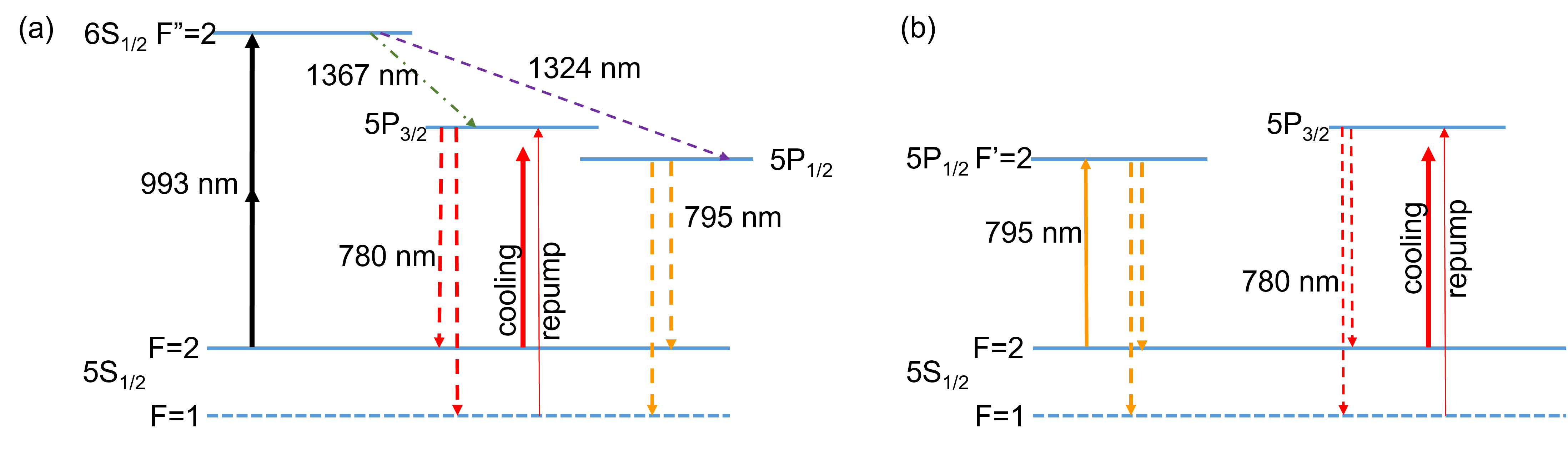}
  
  \caption{Energy level diagrams for  $^{87}$Rb. Cold atoms are prepared in a MOT using cooling and repump beams around 780~nm. Excitation (solid arrows) and decay (dashed arrows) channels are shown. (a) 993 nm light excites  atoms from the 5S$_{1/2}$ F=2 state to the 6S$_{1/2}$ F$''$=2 state using a two-photon process. Atoms decay  to  5S$_{1/2}$ F=1, 2 via the 5P$_{3/2}$ and 5P$_{1/2}$ intermediate levels by emitting 780~nm or 795~nm photons in the lower decay stage. (b) 795 nm light excites atoms from the 5S$_{1/2}$ F=2 state to the 5P$_{1/2}$ F$'$=2 state using a one-photon process. Atoms decay  to  5S$_{1/2}$ F=1, 2  by emitting 795 nm photons. }
  \label{fig1}
\end{figure}

\begin{figure}[!tbp]
  \centering
 \includegraphics[width=\textwidth]{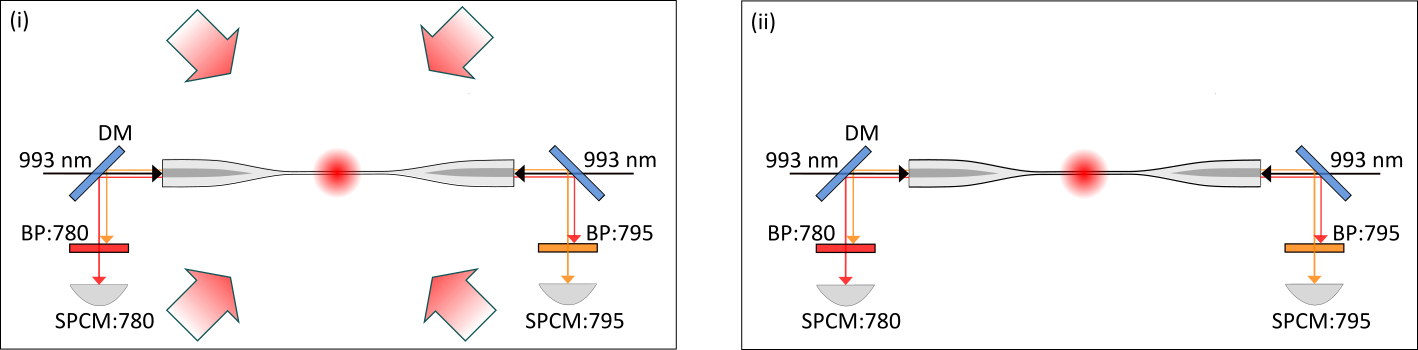}
  \caption{Experimental setups. Two-photon excitation at  993~nm  (\romannum{1}) with  and (\romannum{2}) without  the MOT cooling beams. SPCM: single-photon counting module; BP:bandpass filter; DM: dichroic mirror. The repump beam is kept on at all times.}
  \label{fig2}
\end{figure}

Experiments were performed in two categories: atoms were excited via (i) a two-photon transition or (ii) a one-photon transition, see Fig.~\ref{fig1}.
Resonant and near-resonant excitation of the cold atoms was achieved using a sub-MHz linewidth Ti-Sapphire laser (M Squared Lasers: SolsTis ECD-X). Fluorescence photons during the de-excitation process coupled into the ONF and propagated to the pigtail on either end where they were subsequently detected using single-photon counter modules (SPCM: Excelitas Technologies: SPCM-AQRH-14-FC), see Fig.~\ref{fig2}. Dichroic mirrors and interference filters were used to combine and separate the different wavelengths as per  experimental requirements. Additionally, the fluorescence signal from the cold atom cloud could  be collected in  free-space via  a microscope composed of
two convex lenses, with focal lengths of 125~mm and 250~mm, and was measured using a photomultiplier tube (PMT: Hamamatsu Photonics), see Fig.~\ref{fig3}. Both the lenses and the PMT were kept outside the vacuum chamber. The fluorescence signals were collected for two different configurations as discussed in the following. 

\subsection{Single-frequency, two-photon excitation around the optical nanofibre}\label{TPE exp}
Many experiments involving two-photon processes demand minimizing the one-photon excitation. A larger detuning of the excitation laser frequency with respect to the intermediate level leads to fewer atoms in the intermediate level at the expense of a higher light field intensity requirement.  In single-frequency, two-photon excitation (see Fig.~\ref{fig1}(a)), in general, the laser frequency is very far-detuned from the intermediate levels; additional advantages are the need for only one laser and a more straightforward theoretical description. The subwavelength diameter of an ONF provides very high intensities in the evanescent field at very low powers (e.g. a few nW of propagating light can produce a Rabi frequency of tens of MHz for atoms in the evanescent field \cite{PRA_kumar}), hence providing near-ideal conditions for single-frequency, two-photon excitation to occur.

In this experiment, we used 993~nm light to excite $^{87}$Rb atoms from 5S$_{1/2}$ F=2 $\rightarrow$ 6S$_{1/2}$ F$''$=2 using a two-photon process, see Fig. \ref{fig1}(a). The 993~nm light was obtained from the Ti-Sapphire laser and could be frequency scanned or locked depending on the experiment. Excited atoms in the 6S$_{1/2}$ state spontaneously decayed to the ground state, 5S$_{1/2}$ F=1, 2, via (i) the intermediate state, 5P$_{1/2}$, by emitting a 1324~nm photon followed by a 795~nm photon or (ii) the intermediate state, 5P$_{3/2}$, by emitting a 1367~nm photon followed by a 780~nm photon. The ratio of the transition rates of 6S$_{1/2}$ $\rightarrow$ 5P$_{1/2}$ to 6S$_{1/2}$ $\rightarrow$ 5P$_{3/2}$ is ~0.48 and 5P$_{3/2}$ $\rightarrow$ 5S$_{1/2}$ to 5P$_{1/2}$ $\rightarrow$ 5S$_{1/2}$ is 1.04 \textcolor{red}{\cite{Safronova}}. Hence, it is almost twice as likely that 780~nm photons are emitted in the decay process from 6S$_{1/2}$ to 5S$_{1/2}$ F=1, 2 than 795~nm photons.    

\subsubsection{Cooling beams on:}
In the first configuration, 993~nm excitation light was sent through the ONF and fluorescence measurements were made in the presence of the MOT i.e. with cooling and repump beams on. The 993~nm light was frequency-scanned across the 5S$_{1/2}$ F=2 $\rightarrow$ 6S$_{1/2}$ F$''$=2 resonance ($\sim$301.77791 THz).   The fluorescence signals at 795~nm and 780~nm coupled into the ONF and were measured at the SPCMs, see Fig. \ref{fig2}(\romannum{1}). In this scenario, fluorescence at 780~nm had contributions from both the excitation process to the 6S$_{1/2}$ F$''$=2 state (via one of the decay channels) and the MOT beams exciting atoms to the 5P$_{3/2}$ F$'$ states. In contrast, detected fluorescence photons at 795~nm were solely due to the 6S$_{1/2}$ F$''$=2 state excitation, see the level diagram in Fig. \ref{fig1}(a).

\subsubsection{Cooling beams off:}\label{wo cooling}

 Here, the MOT cooling beams were turned off during the fluorescence measurement time window, see Fig.~\ref{fig2}(\romannum{2}). In this configuration, the fluorescence signals at 795 nm and 780 nm were mainly from the 6S$_{1/2}$ F$''$=2 decay channels.  Though some repump signal could  contribute we assumed this was negligible. The experimental sequence was as follows: 993 nm light was sent through the ONF and its frequency was locked near or on resonance. The cold atom cloud was initially prepared in the MOT and this was followed by a polarisation gradient cooling phase (bringing the cooling  power per beam from 11~mW/cm$^2$ to 2.7~mW/cm$^2$ and the cooling beam detuning from -14MHz to -30 MHz)  for 5~ms. Subsequently, the cooling beams were turned off before the measurements.  The fluorescence signal was collected during the first 700 $\mu$sec and averaged over 20 cycles. A full spectrum was obtained by scanning the locking point of the 993~nm laser in  steps of $\sim$2~MHz across the resonance and repeating the  procedure as described above for each frequency step.  

\subsection{One-photon excitation in a cold atom cloud}
To better understand the observed features in the two-photon excitation experiments, we also performed one-photon excitation experiments as follows.  Cold atoms in the MOT were excited from the 5S$_{1/2}$ F=2 state to the  5P$_{1/2}$ F$'$=2 state via 795~nm light ($\sim$ 377.10519~THz), see Fig.~\ref{fig1}(b). Since the excitation and the fluorescence light were of the same wavelength, the detection scheme used in Section \ref{TPE exp}, i.e. sending the excitation light through the ONF and  detecting the fluorescence signal using a SPCM, was not suitable. Instead, here we used a free-space beam at 795 nm, focussed on the cold atom cloud via a convex lens (Thorlabs:LB1779-B, f=30 cm). The excitation light at 795~nm was derived from the Ti-Sapphire laser and could  be scanned or locked at a fixed frequency. The experiments were conducted using the following  two configurations. 
\subsubsection{Cold atoms around the ONF:}

\begin{figure}[!tbp]
  \centering
 \includegraphics[width=\textwidth]{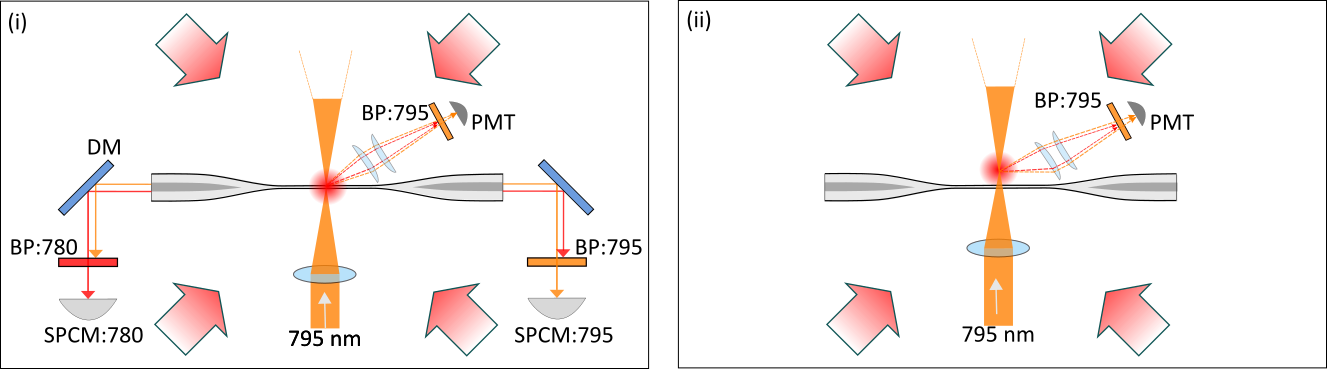}
  \caption{Experimental setups. One-photon excitation at 795~nm  with the cold atom cloud (\romannum{1}) around the ONF  and  (\romannum{2}) removed from the ONF. SPCM: single-photon counting module; PMT: photomultiplier tube; BP:bandpass filter; DM: dichroic mirror.} 
  \label{fig3}
\end{figure}

The MOT was formed around the ONF while the focussed excitation beam at 795~nm, transverse to the ONF, was sent through the atomic cloud, see Fig.~\ref{fig3}(\romannum{1}). As the frequency of the 795~nm was scanned across the resonance, the fluorescence signals at 795~nm and 780~nm were measured using SPCMs and a PMT.

\subsubsection{Cold atoms removed from the ONF:}

In this case, we performed the one-photon excitation at 795~nm in a MOT when the cold atom cloud was at some distance from the ONF, see Fig.~\ref{fig3}(\romannum{2}). This was achieved by changing the current in one of the MOT anti-Helmholtz coils, thereby shifting the zero of the magnetic field, hence the trap position. The excitation beam was realigned so that it passed through the atom cloud. The fluorescence signal in this configuration could only be measured by the PMT since the ONF was too far from the cold atom cloud for fluorescence coupling into the fibre.

\begin{figure}[!tbp]
  \centering 
 \includegraphics[width=0.48\textwidth]{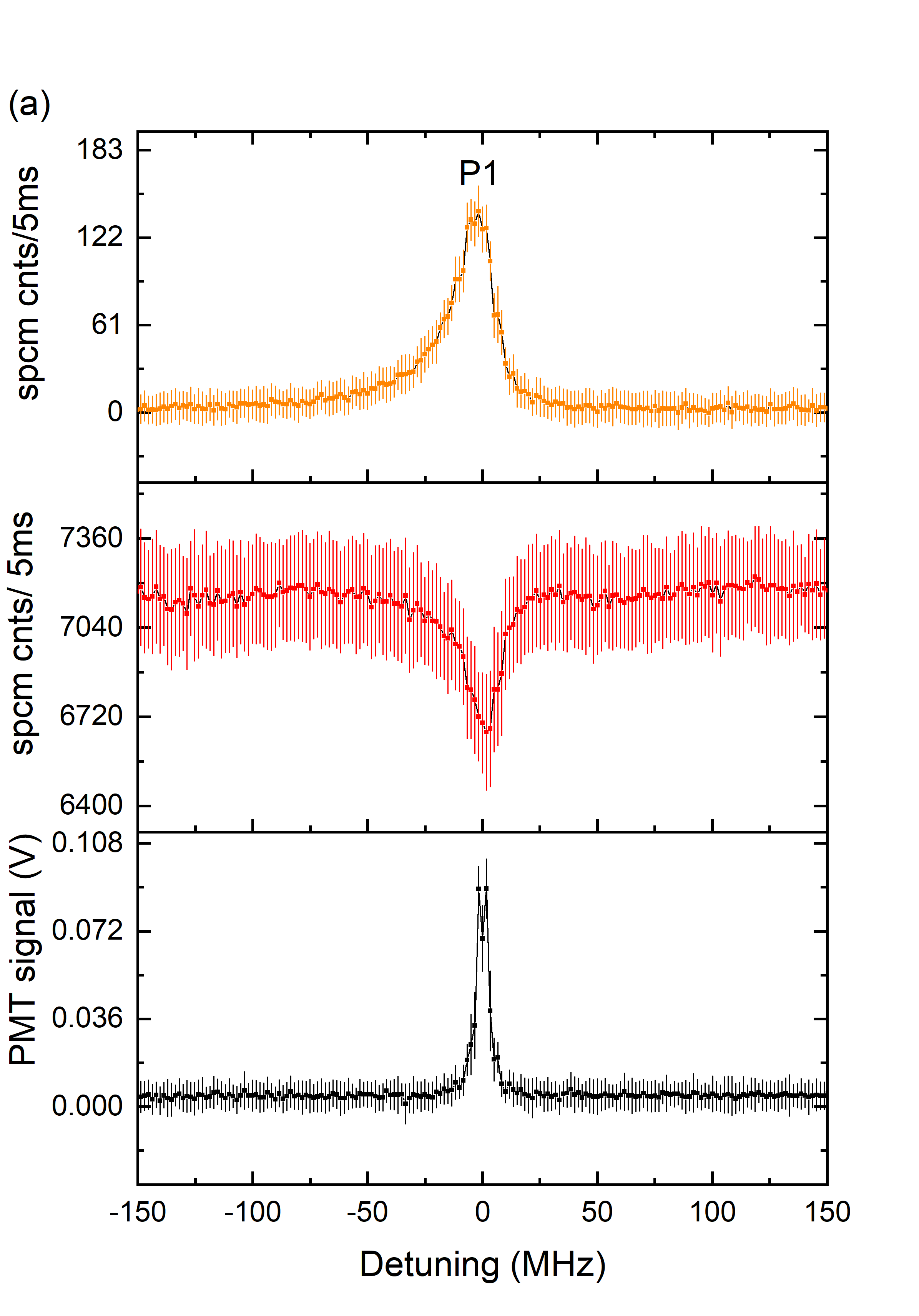}
  \hfill
 \includegraphics[width=0.48\textwidth]{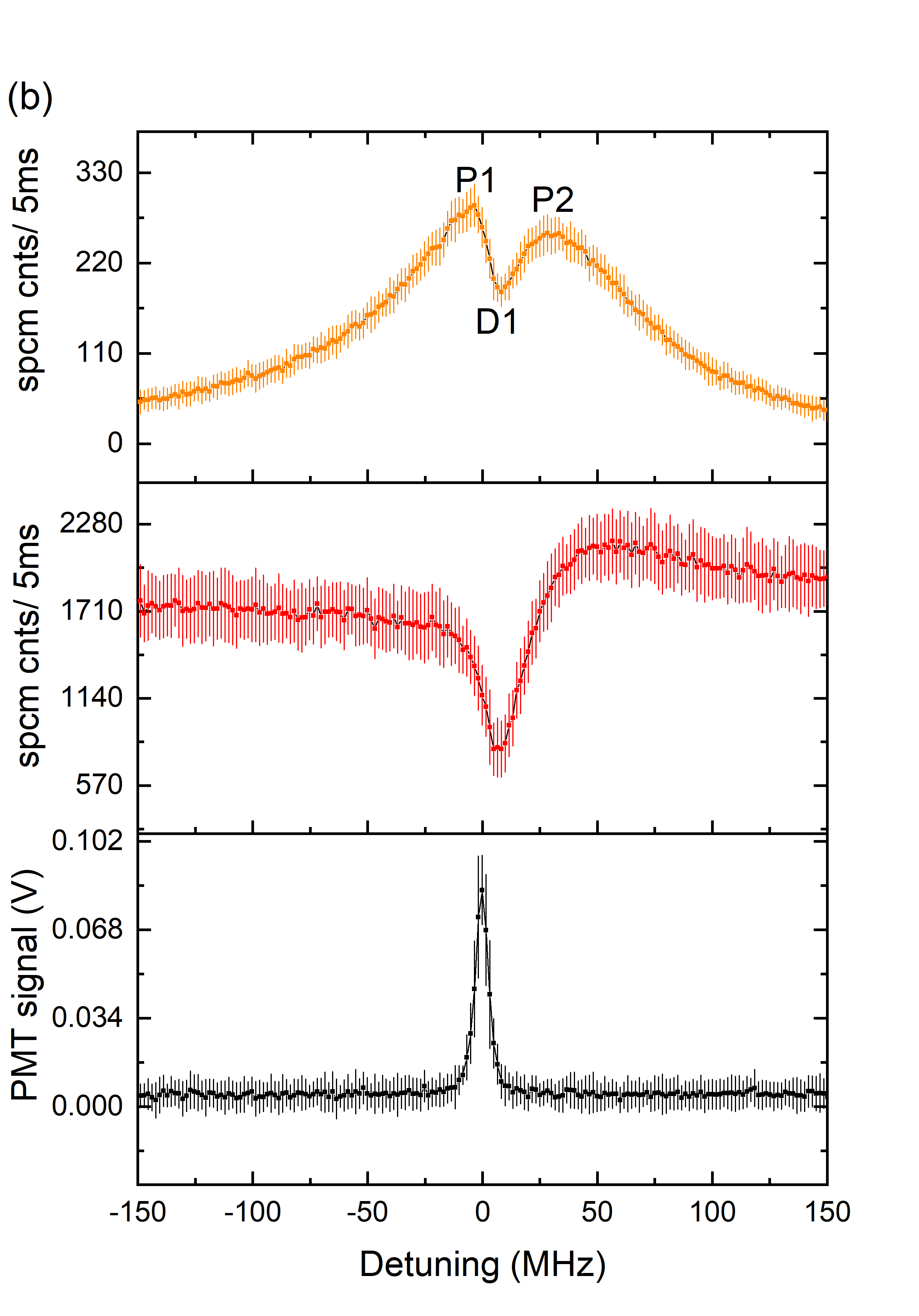}
  \caption{Excitation light at 993 nm, going through the ONF in both directions, is scanned across the 5S$_{1/2}$ F=2 $\rightarrow$ 6S$_{1/2}$ F$''$=2 transition via two-photon excitation.  Total power of 993~nm excitation light in the ONF is (a) 0.1~mW and (b) 2.45~mW. From top to bottom: 795~nm fluorescence signal coupled into the ONF, 780~nm fluorescence signal coupled into the ONF, spectroscopy signal from a Rb vapour cell recorded on a PMT for the frequency reference. Typical laser power used for the vapour cell spectroscopy was $\sim$ 100 mW.}
  \label{fig4}
\end{figure}

\section{Experimental Results}

\subsection{Single-frequency, two-photon excitation}

Let us first discuss the results obtained for a cloud of cold atoms in a magneto-optical trap surrounding the ONF.  For a low power (0.1~mW) of the 993 nm excitation light propagating through the ONF, as the laser is tuned on-resonance with the 5S$_{1/2}$ F=2 $\rightarrow$ 6S$_{1/2}$ F$''$=2 two-photon resonance condition, we observe a peak (P1) in the 795~nm fluorescence and a corresponding dip in the 780~nm fluorescence signal coupled into the nanofibre, see top and middle panels of Fig. \ref{fig4}(a). Detuning is defined as  twice the laser frequency minus the atomic transition frequency for 5S$_{1/2}$ F=2 $\rightarrow$ 6S$_{1/2}$ F$''$=2. With an increase in the 993 nm power to 2.45~mW, the 795~nm fluorescence signal shows a dip, D1, and the appearance of a second peak, P2, as shown in the top panel of Fig. \ref{fig4}(b). P2 becomes more pronounced with an increase in the 993~nm power; the two peaks, P1 and P2, also move further apart, i.e., the dip appears to become broader. The dip in the 780~nm signal is at exactly the same frequency as the dip in the 795~nm signal, see the top and  middle panels of Fig.~\ref{fig4}(b). The dip appears at a slightly positive detuning (by 1 to 7~MHz) (see bottom panel). Frequency calibration was obtained via simultaneous measurement using a wavemeter (HighFinesse GmbH WS8-2) in conjunction with spectroscopy performed in a Rb vapour cell with natural isotope abundance. The vapour cell was heated to $\sim$100$^{\circ}$C. Counterpropagating 993~nm laser beams passing through the cell cancelled the first-order Doppler shift and provided us with a Doppler-free spectroscopy peak as a precise frequency reference \cite{Nieddu:19}. The full-width-at-half-maximum of the vapour cell signal is $\sim$6 MHz, see bottom panels of Fig.~\ref{fig4}. The fluorescence signal obtained from the cold atoms near the ONF was typically much broader ($\sim$ 20 MHz) than the vapour cell signal, which we attribute to surface effects from the ONF \cite{Nayak2008split,2009spectraldistribution}. We also observed that the splitting was more pronounced when the cloud was less dense (data  not shown).

\begin{figure}[!tbp]
\includegraphics[width=\textwidth,keepaspectratio,]{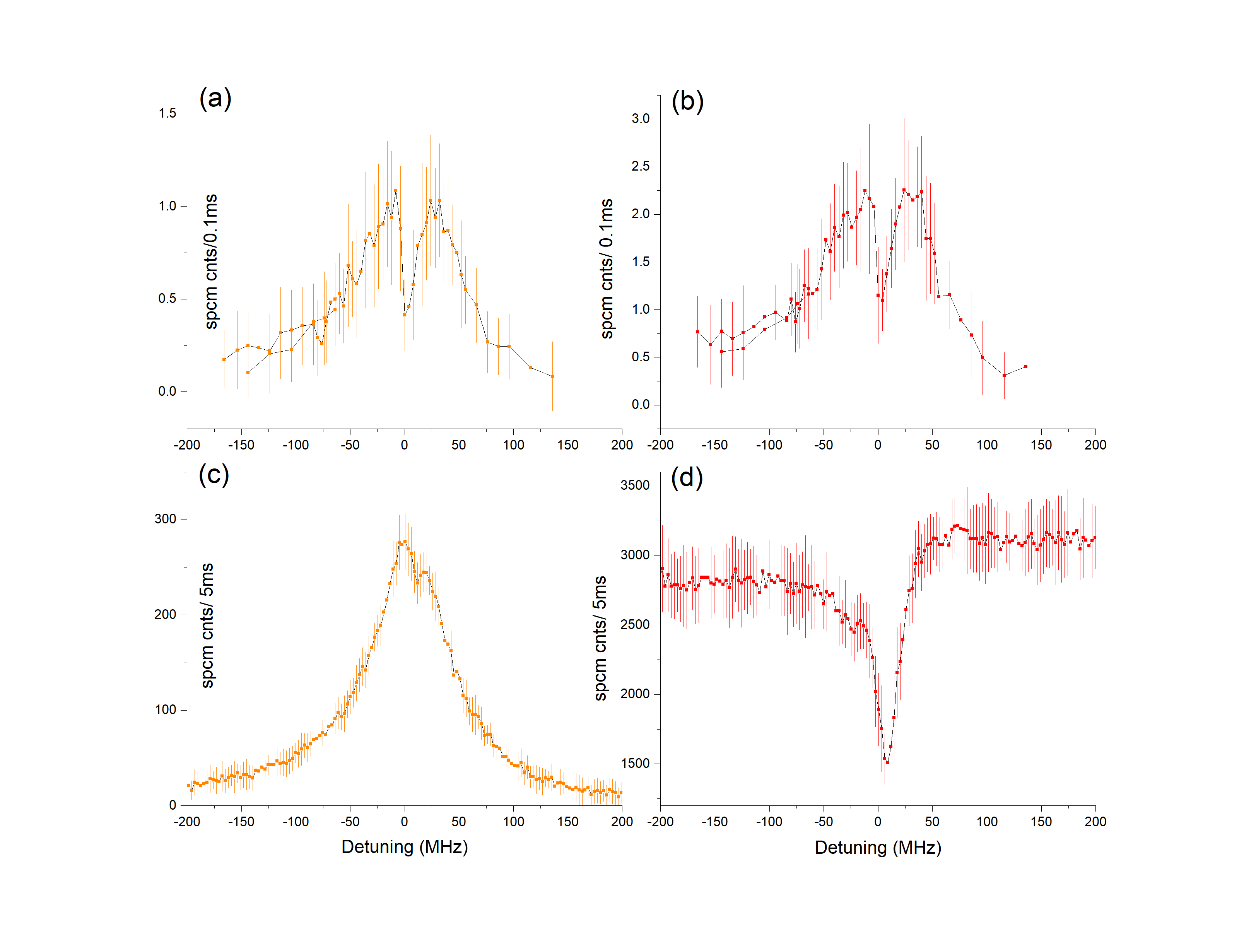} 
\centering
 \caption{Fluorescence signal during two-photon excitation at 993 nm. Comparison of 795 nm (left plots in orange) and 780 nm (right plots in red) fluorescence signals in the absence (a, b) and in the presence (c, d) of cooling light. Excitation light through the ONF was kept at 2 mW.}\label{fig5}
\end{figure}

\begin{figure}[!tbp]
\includegraphics[width=0.8\textwidth,keepaspectratio,]{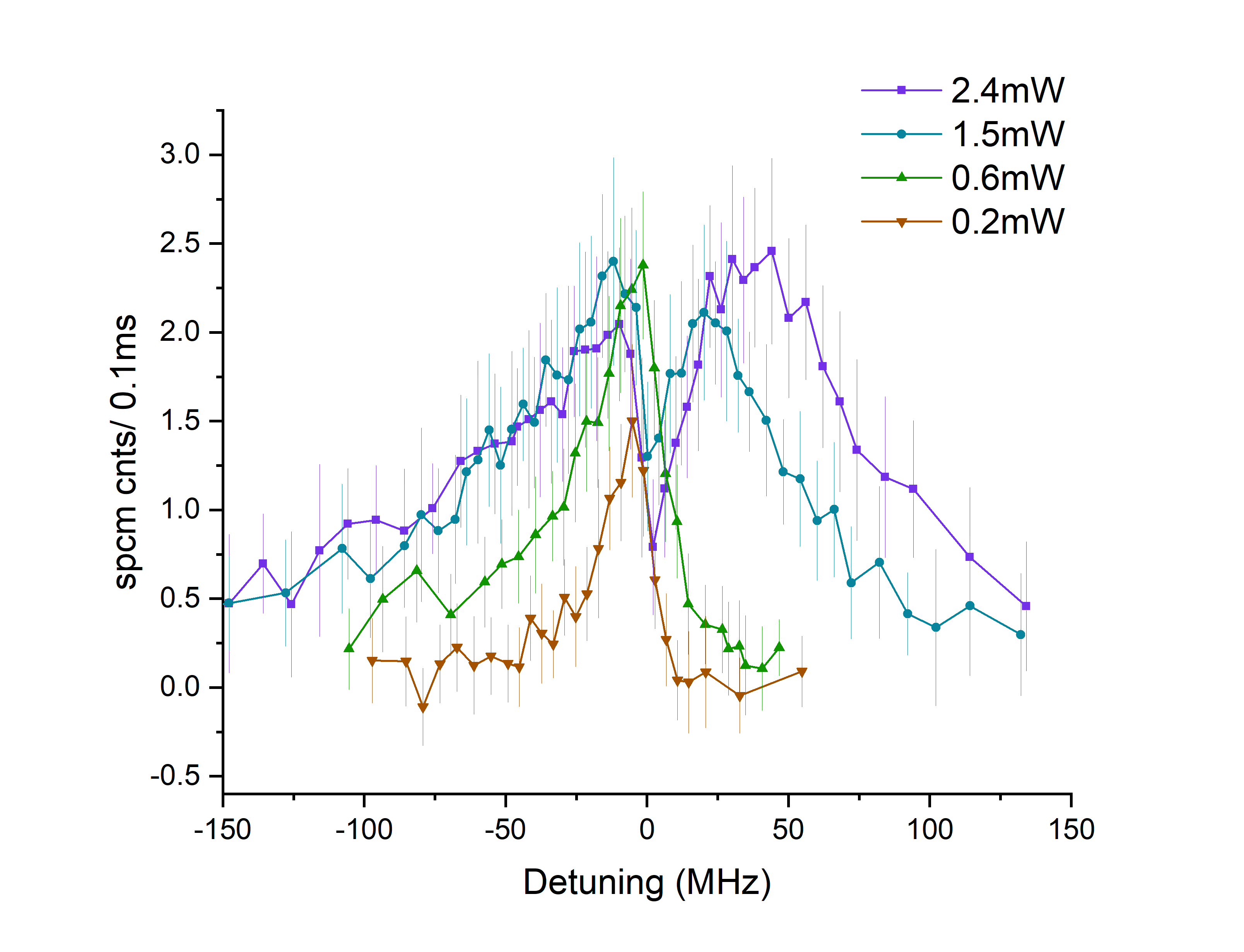} 
\centering
 \caption{Fluorescence at 795 nm from the cold atoms near the ONF during two-photon excitation in the absence of the cooling beams. Signals are compared for  different excitation powers.}\label{fig6}
\end{figure}

To simplify the system being studied, we next ran a series of tests with the cooling beams at 780~nm switched off during data acquisition and 2~mW of 993~nm excitation power going through the ONF.  For the experiments presented in Fig. \ref{fig4}, the cooling beams were on at all times.  By running the experiments in the absence of cooling beams, we could eliminate the possibility of Autler-Townes splitting being the cause for the observed split profiles.   In the absence of the cooling beams, the observed fluorescence signals at 780~nm and 795~nm show similar trends with different amplitudes, see Fig.   \ref{fig5}(a) and (b). This is expected since both  signals are from the de-excitation of the 6S$_{1/2}$ F$''$=2 state to the 5S$_{1/2}$ F=1, 2 states albeit with different transition probabilities (as discussed in Section \ref{TPE exp}). For comparison, fluorescence signals at 780~nm and 795~nm in a MOT, i.e., with the cooling beams on, are also  plotted, see  Fig. \ref{fig5}(c) and (d). Comparing Fig. \ref{fig5}(a) and (c), we see that the splitting is present in both the cases though it is more pronounced and  the dip is deeper in the absence of cooling beams compared with the signals obtained with cooling beams.

Finally, in Fig. \ref{fig6}, the fluorescence signal at 795~nm for different 993~nm powers, in the absence of cooling beams, is plotted. Similar to the results presented in the top panels of Fig. \ref{fig4}(a) and (b), we observe a two-peak profile and the dip becomes more pronounced as the 993~nm power is increased.

\begin{figure}[!tbp]
  \centering
 \includegraphics[width=0.48\textwidth]{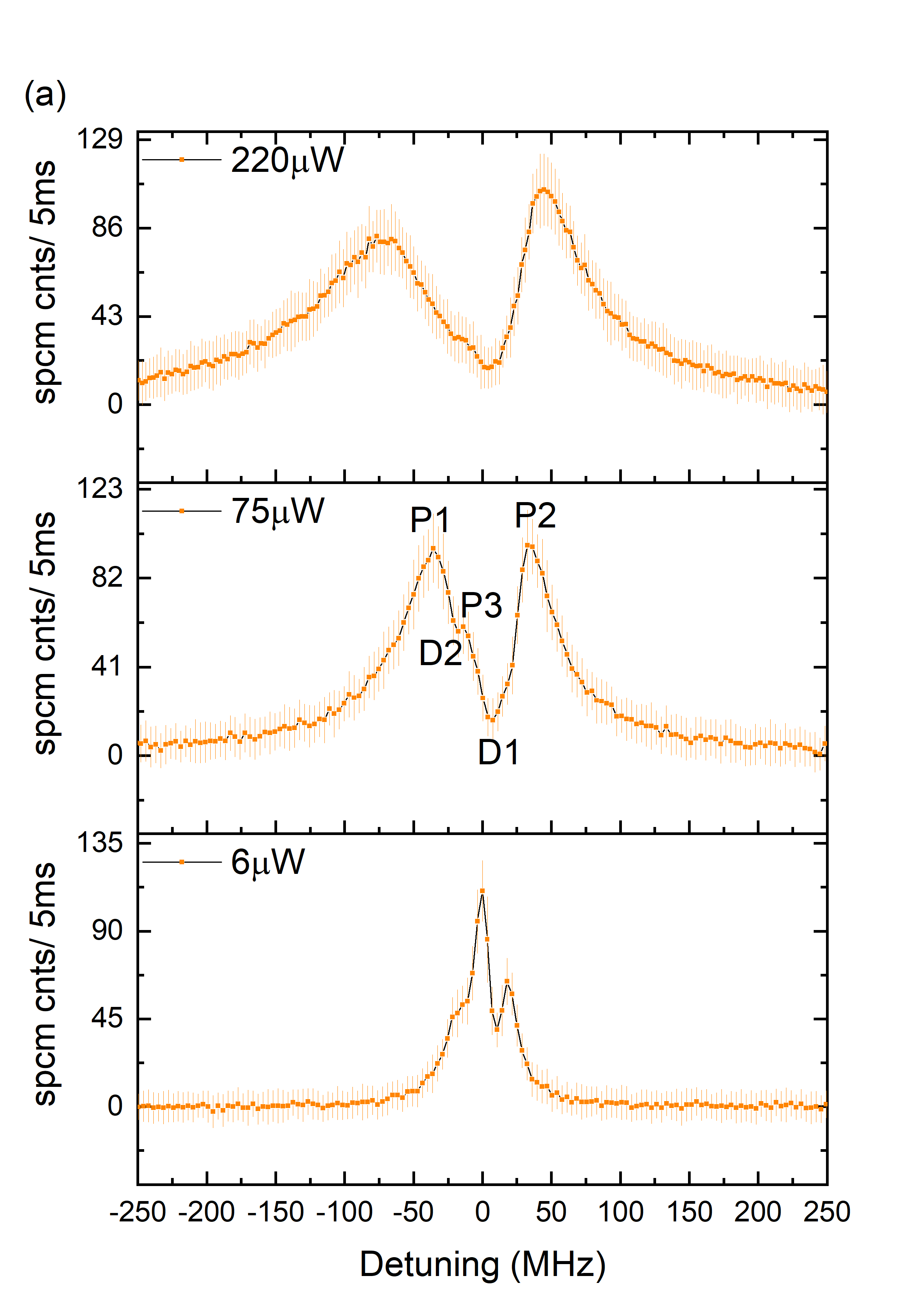}
  \hfill
\includegraphics[width=0.48\textwidth]{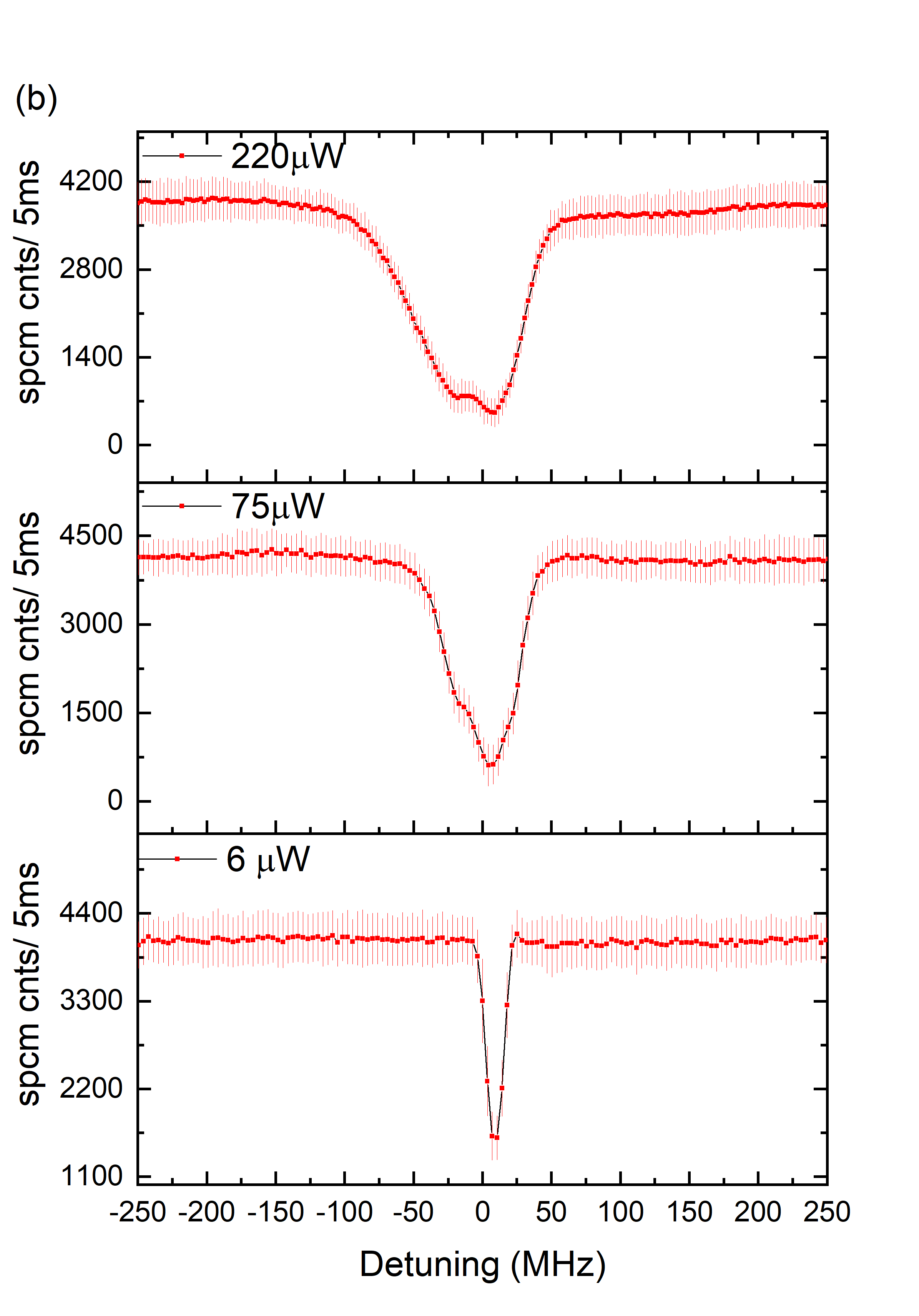}
  \caption{Fluorescence signals from the MOT during one-photon excitation at 795 nm. (a) 795 nm fluorescence signal for different excitation powers of 795 nm. (b) 780 nm fluorescence signal for different excitation powers of 795 nm.} \label{fig7}
\end{figure}

\subsection{One-photon excitation}
\begin{figure}[!tbp]
  \centering
  \includegraphics[width=0.48\textwidth]{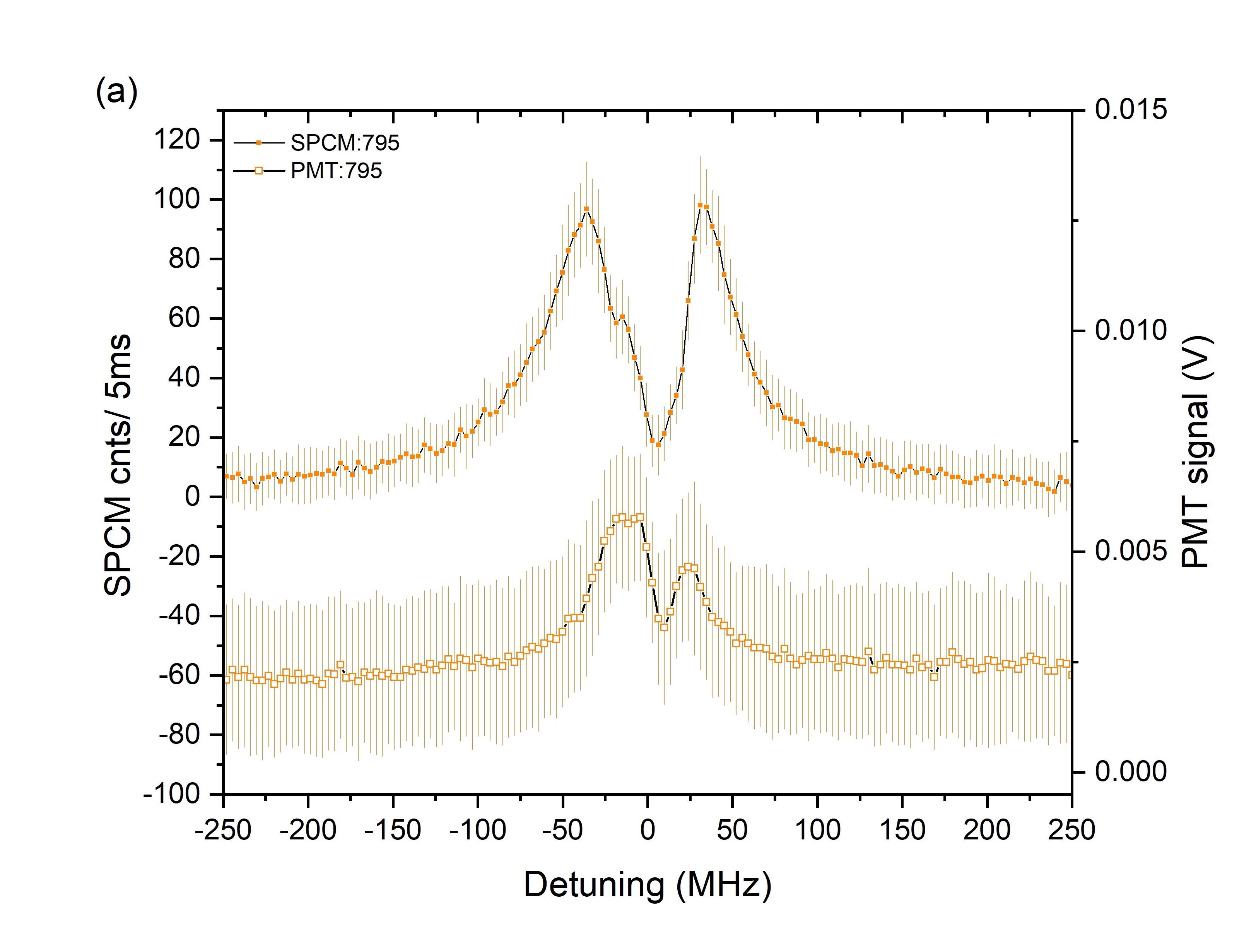}
  \hfill
 \includegraphics[width=0.48\textwidth]{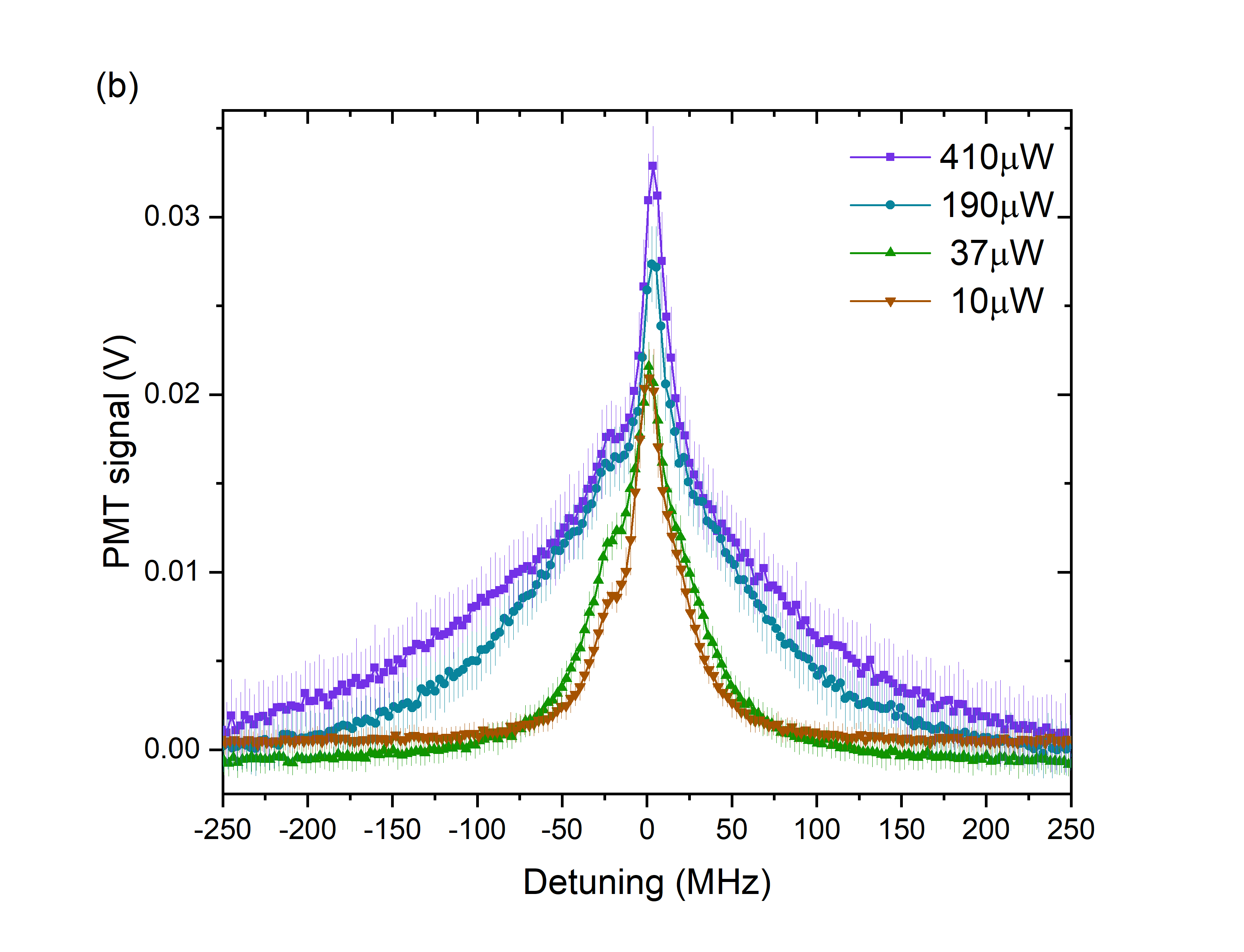}
    \caption{(a) One-photon excitation signal when the cold atom cloud overlaps the ONF. Comparison of 795~nm signals collected from the PMT and the SPCM are shown by empty (bottom curve) and solid (top curve) circle data points, respectively. (b) One-photon excitation when the cold atom cloud is far from the ONF. Fluorescence signal at 795 nm on the PMT is plotted for different excitation powers.}
		\label{fig8}
  
\end{figure}

The intriguing two-peak profiles observed in the two-photon excitation experiments were investigated by comparing the results with two different, one-photon excitation configurations,  discussed in the following.   The first configuration was when the atom cloud surrounded the optical nanofibre.  Atoms in the MOT were excited from 5S$_{1/2}$ F=2 to 5P$_{1/2}$ F$'$=2  via a focussed 795~nm light in free-space. Emitted fluorescence photons at 795~nm and 780~nm, which coupled into the ONF, were detected via the SPCM, see Fig. \ref{fig3}(\romannum{1}).  The observed spectra are shown in Fig. ~\ref{fig7}(a) and (b) where  detuning is defined as the laser frequency minus the atomic transition frequency for  5S$_{1/2}$ F=2 $\rightarrow$ 5P$_{1/2}$ F$'$=2. The fluorescence at 795~nm, see Fig.~\ref{fig7}(a),  exhibits two dominant peaks (P1 and P2) and a dip (D1)  as the 795~nm excitation laser was detuned from resonance. The spectra also show a small dip (D2) and a peak (P3). Peak P2 becomes more prominent as the 795~nm excitation power is increased and P1, P2 and D1 become broader, see  Fig.~\ref{fig7}(a). Fluorescence was also measured at the PMT without being coupled to the ONF, see Fig. \ref{fig3}(\romannum{1}) for the experimental setup. When we compare the fluorescence signals measured at the PMT and the SPCM, see Fig. \ref{fig8}(a), we observe that the PMT signal is narrower and shows a dip at the same position as the SPCM signal. These observations show that, similar to the two-photon excitation scenario, here also the two peaks, P1 and P2, move apart or the dip, D1, becomes broader as the excitation power is increased. 

The second one-photon configuration we considered was when the cold atom cloud was distant from the ONF, see Fig. \ref{fig3}(\romannum{2}) for the setup, to ensure that there was no coupling of the fluorescence signal into the nanofibre.  In this case, the fluorescence signal at 795~nm as a function of the excitation laser detuning from  5S$_{1/2}$ F=2 $\rightarrow$ 5P$_{1/2}$ F$'$=2 transition, see Fig.~\ref{fig1}(b),  was collected by the PMT. In Fig.~\ref{fig8} (b), the fluorescence signal at 795~nm while varying the 795~nm excitation laser frequency is plotted for different excitation powers. Even at higher powers (410~$\mu$W) a second peak does not appear in the spectra. Additionally, the observed  fluorescence profile is not Lorentzian. 

\begin{figure}[!tbp]
  \centering
 {\includegraphics[width=0.48\textwidth]{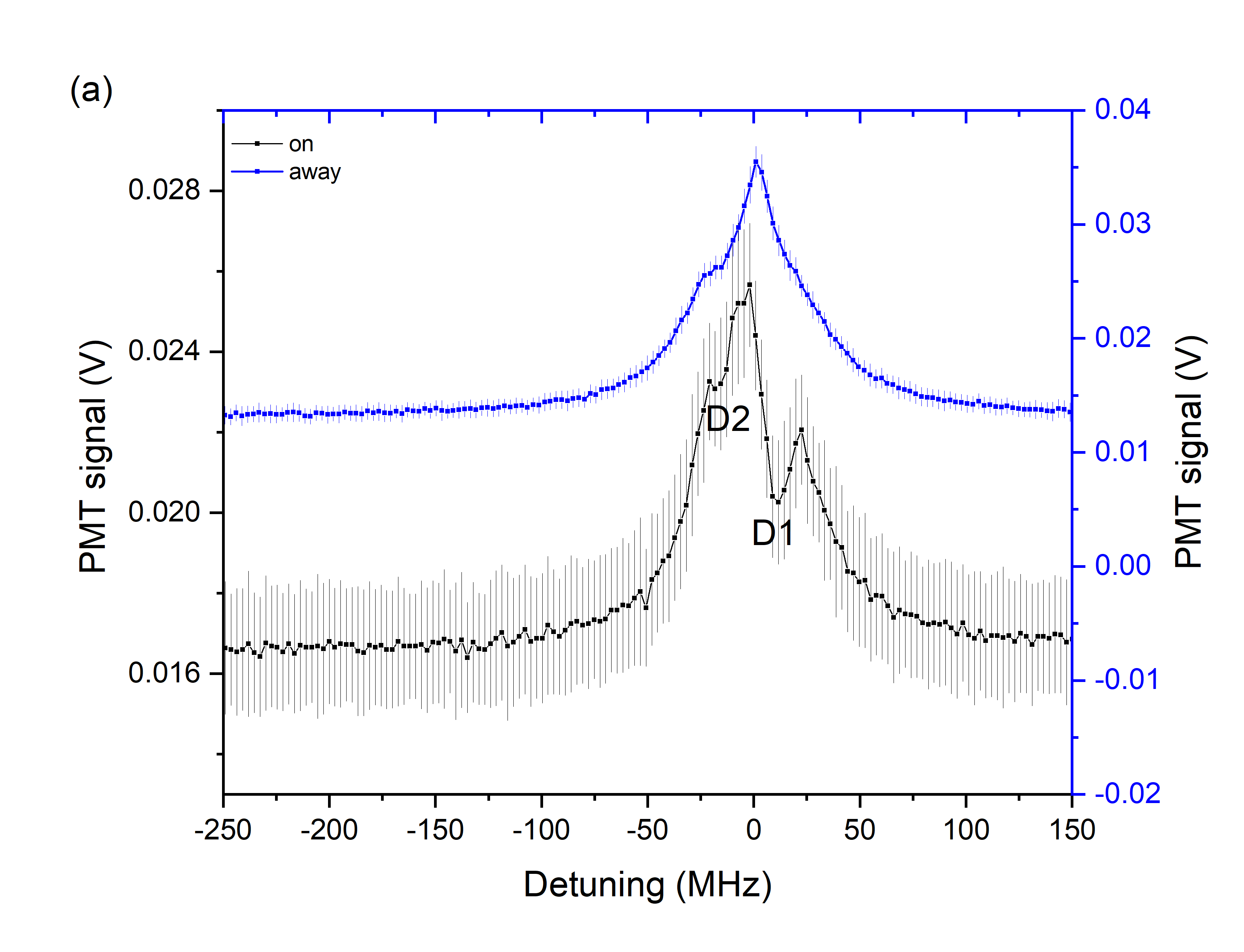}}
  \hfill
 {\includegraphics[width=0.48\textwidth]{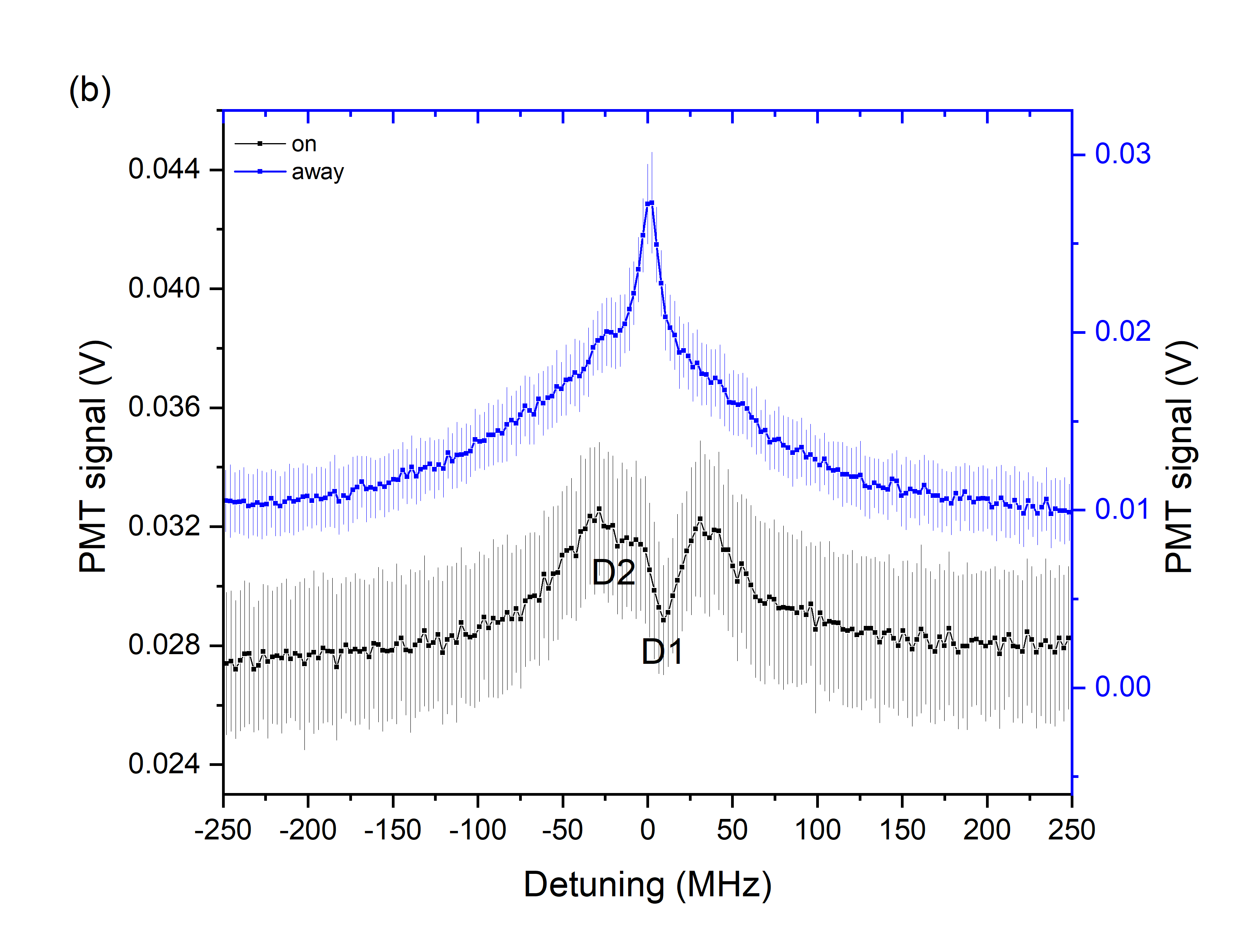}}
  \caption{Comparison of 795 nm fluorescence signal at the PMT during one-photon excitation. Data in blue show the signal when MOT is formed away from the ONF and data in black show when MOT is on the ONF. Excitation powers are $\sim$37 $\mu$W and $\sim$150 $\mu$W for (a) and (b), respectively. }
  \label{fig9}
\end{figure}

Finally, to compare the observed spectra  when the atom cloud overlaps the ONF and when the cloud is removed from the ONF, we measured the 795~nm fluorescence signals for the same excitation powers. The results are plotted in Fig.~\ref{fig9} for two different excitation powers and show the striking contrast between the two cases.  Notably, the two-peak profile is present only when both the cold atom cloud and the focussed, strong excitation beam are at the ONF.    In both these cases, the 795~nm excitation light was focussed at the atom cloud via the lens.  

\section{Discussion}
Results show that, for both one- and two-photon excitation processes near an ONF, high intensities of the evanescent field produce a two-peak profile in the fluorescence signal as a function of laser detuning. The similarity of the profiles in the two cases suggests that the origin is not linked with any process specific to the one- or two-photon excitation phenomena. To better understand the observed structure, we considered the feasibility of the shape arising from certain phenomena to hypothesise on the exact origin. 

As a first line of investigation, we  considered whether the lineshape could arise from interference.  Subnatural linewidth spectral lines have already been investigated theoretically and experimentally both in V-type and $\Lambda$-type atomic systems in the presence of  two laser fields \cite{PhysRevLett.Zibrov, HOPKINS1997185}. Quantum interference between the two transition pathways plays a crucial role in producing such narrow spectral features. Interference can also be present when monochromatic light couples a single atomic ground state to two closely spaced excited states by parallel dipole moments. At an appropriate frequency of the excitation laser field, fluorescence from the excited levels can be eliminated \cite{Cardimona_1982}. However, the condition of parallel dipole moments cannot be fulfilled when considering two closely spaced Zeeman sublevels of a particular hyperfine state, hence making it nontrivial to achieve. It has also been proposed that anisotropy of the vacuum field could provide the necessary condition for  quantum interference among closely lying states \cite{anisotropy_agarwal}. Since the two observed peaks are far-separated even at zero magnetic field, we rule out this possibility. In summary, interference cannot give a satisfactory answer for the origin of the two-peak profile.  

Another possibility we considered was that of Autler-Townes splitting \cite{PRA_kumar} caused by the strong laser cooling beams; however, we have also ruled this out because the splitting is present even in the absence of the cooling beams and theoretical models do not support this possibility. 

We propose that the actual phenomenon is far simpler than initially assumed. The observed spectra can be explained by considering the pushing effect of the excitation light,  i.e., 993~nm or 795~nm laser light, on the atoms. As the light frequency approaches the atomic resonance, the photon absorption rate increases. Atoms in the highest intensity region are pushed away maximally due to momentum kicks given by the photons. This leads to a reduced atom density locally and, thereby, reduced fluorescence. In addition, we also need to consider that the atomic resonance or the resonant transition frequency in a MOT is modified due to the ac Stark effect of the cooling beams. These new resonance frequencies will be referred henceforth as the modified resonances. The experiments in the absence of cooling beams also nitially  relied on having a MOT, i.e., cooling light, in the presence  of a frequency-locked excitation beam passing through the ONF (see Subsection \ref{wo cooling}). Due to the aforementioned pushing effect, the initial atom number near the ONF depends on the excitation frequency and is not the same for each frequency step  during the scan. This leads to low fluorescence whenever the frequency of the excitation light matches to the modified resonances,  even though fluorescence was collected in the absence of the cooling beams.  In the following, to understand the frequency values or positions of the modified resonances, we consider the ac Stark effect of the cooling beams on both the ground (5S$_{1/2}$ F=2) and the excited  (5P$_{3/2}$ F$'$=3) states of an $^{87}$Rb atom. The new energy eigenstates can be calculated using the dressed atom picture \cite{1998atomPhotonCohen}.
 
To compare with the bare atom energy states, we consider the specific case of a red-detuned light field, such as the MOT cooling beams in our experiments. In this case, both the ground and the excited states split into two dressed states. The two dressed ground states are shifted in frequency by $\Delta_{1g} = - (\Omega_c- |\delta_c|)/2$ and $\Delta_{2g} = (\Omega_c+|\delta_c|)/2$  w.r.t. the bare atom ground state. Similarly, the excited dressed states are shifted by $\Delta_{1e} = (\Omega_c- |\delta_c|)/2$ and $\Delta_{2e} = -(\Omega_c+|\delta_c|)/2$  w.r.t. the bare atom excited state.
Here, $\delta_c$ is the cooling beam detuning, $\Omega_c$ = $\sqrt{(\delta_c)^{2} + (\Omega_{c0})^{2})}$, and $\Omega_{c0}$ is the on-resonant Rabi frequency due to the cooling beams.
In the limit of high detunings, the ground state shifts can  be approximated as $\Delta_{1g} =  -{\Omega_{c0}^{2}}/(4|\delta_c|)$ and  $\Delta_{2g} = - \delta_c + {\Omega_{c0}^{2}}/(4|\delta_c|)$. Dressed atom spectroscopy has been investigated in detail by several groups earlier, see \cite{Mitsunaga:96,Wu}  

 Thus, as the frequency of the excitation laser is scanned (during the one- or two-photon excitation) within a MOT, there are two possible resonances, namely $\Delta_{1g}$ $\rightarrow$ excited state ($\Delta_{1}$ resonance) and $\Delta_{2g}$ $\rightarrow$ excited state ($\Delta_{2}$ resonance). In the one-photon excitation, see Fig.~\ref{fig7}(a), we observe that the larger (D1) and smaller dips (D2) in the 795~nm fluorescence signal are approximately at the frequencies of the $\Delta_{1}$ and $\Delta_{2}$ resonances. From this, we speculate that the dip, which is shifted by a few MHz from the 5S$_{1/2}$ F=2 $\rightarrow$ 5P$_{1/2}$ F$'$=2 resonance, is due to the pushing of the atoms occurring at the $\Delta_{1}$ resonance. The pushing effect at the $\Delta_{2}$ resonance appears much smaller than that at $\Delta_{1}$.

The pushing effect is much more significant when atoms are near the ONF, probably since the field intensity is much higher and there is much more light scattering in this region. Both for the one- and two-photon excitation processes, the broad fluorescence profile with a relatively narrower dip is due to a varying power broadening and a varying pushing effect in the evanescent field due to its exponential profile. 
We emphasise that the two-peak profile is observed only for the excitation of atoms near a fibre surface. When the detection is also via the nanofibre, the dip is more pronounced and broader. This may be due to higher scattering and pushing near the fibre surface and probing it efficiently via the nanofibre itself. In contrast, when atoms are far away from the ONF, detection is done using a PMT which gives the collective signal from the MOT. We observe a small dip at $\sim$ -14MHz but the positive side dip (D1) is not visible, see the PMT signals in Fig. \ref{fig9}(a) and (b).

During these studies, we came to know that a similar spectrum has been observed with Cs atoms in a MOT near an optical nanofibre \cite{Nayak2008split,kpNayak} where the authors did the following experiments. For a single-atom condition near the ONF, a  probe laser in a travelling wave configuration was focussed perpendicular to the fibre axis. Fluorescence collected from the ONF showed a sharp dip (narrower than the natural linewidth) as the probe frequency was scanned across the resonance. The authors speculated that the  effect arose from the shifting of the MOT centre due to  pushing by the probe beam, thereby changing the local density of the atom cloud. Another hypothesis for the observed spectrum was quantum interference and the result was attributed to atom trapping with motional quantisation \cite{spie_Hakuta}. Our results support the former hypothesis though our results are neither limited to the single-atom condition nor one-photon excitation. Additionally, our studies show that this effect is present when light propagates through the nanofibre either in a travelling wave or counter-propagating configuration.

\section{Conclusion}
We have experimentally investigated the origin of the two-peak profile of a fluorescence signal when  intense, single-frequency two-photon excitation light interacts with cold atoms near an optical nanofibre. To understand the origin of this profile, we performed experiments using one-photon excitation for atoms both near and far from the nanofibre. Observations indicate that a similar profile is present when one-photon excitation is performed near the ONF. We speculate that, at higher excitation powers, resonance scattering induced pushing near the nanofibre becomes the dominant effect. This sharply depletes the atom number density, giving rise to a dip in the fluorescence. The frequency position of the resonance is modified due to the laser cooling beams; hence, the dip is slightly shifted from the bare atom resonance. This hypothesis also explains the similarity of profiles obtained in the presence and absence of cooling beams during two-photon excitation. This is due to pushing, or loss, of the atoms in a MOT, resulting in a lower initial atom number in the molasses phase.  The dip is also observed in the molasses phase. These observations are crucial for  near-resonant, high-intensity experiments based on the cold atom-nanofibre platform in order to better understand the spectral profiles obtained.

\ack
This work was supported by the Okinawa Institute of Science and Technology Graduate University and JSPS KAKENHI (Grant-in-Aid for Scientific Research (C)) 19K05316. S.N.C. acknowledges support by Investments for the Future from LabEx PALM (ANR-10-LABX-0039-PALM).  The authors thank A. M. Akulshin, J. L. Everett, H.-H. Jen, J. Mompart, K. P. Nayak and J. Robert for insightful discussions, and T. Ray and K. Subramonian Rajasree for their early studies that inspired this work.

\clearpage

\section*{References}

\bibliographystyle{iopart-num}

\bibliography{iopart-num}

\end{document}